\documentclass[preprint, 12pt]{elsarticle}
\usepackage{amsmath,amssymb,amsfonts,amsbsy}
\usepackage[nodots]{numcompress}
\usepackage{aas_macros}
\usepackage{mathrsfs}
\usepackage{url}
\usepackage{multirow}
\usepackage{xcolor}
\usepackage{color}
\usepackage{ulem}
\usepackage{enumerate}
\usepackage[linkcolor=blue,citecolor=blue,backref=page]{hyperref} 

\newcommand{\dif}{\mathrm{d}}

\begin{document}
\begin{frontmatter}
\title{Constraints on Dark Matter Annihilation/Decay from the Isotropic Gamma-Ray Background}

\author{Wei Liu}
\author{Xiao-Jun Bi}
\author{Su-Jie Lin}
\author{Peng-Fei Yin}

\address{Key Laboratory of Particle Astrophysics,
Institute of High Energy Physics, Chinese Academy of Sciences,
Beijing 100049, China}

\begin{abstract}
In this work we study the constraints on dark matter (DM) annihilation/decay from the Fermi-LAT Isotropic Gamma-Ray Background (IGRB) observation. We consider the contributions from both extragalactic and galactic DM components. For DM annihilation, the evolutions of extragalactic DM halos are taken into account.
We find that the IGRB constraints under some DM subhalo models can be comparable to those derived from the observations of dwarf spheroidal galaxies. We also use the IGRB results to constrain the parameter regions accounting for the latest AMS-02 electron-positron anomaly. We find that the majority of DM annihilation/decay channels are strongly disfavored by the latest Fermi-LAT IGRB observation; only DM annihilation/decay to $\mu^+\mu^-$ may be valid.
\end{abstract}
\end{frontmatter}

\section{Introduction}
\label{INTRO}
From the numerous observations of the astrophysics and the cosmology, it is well confirmed that the dark matter (DM) constitutes about $84\%$ of the total matter in the universe \citep{2015arXiv150201589P}. Despite of its proverbial existence, we still have a poor understanding on its microscopic properties. In many new physics models, a kind of weakly interacting massive particles(WIMPs) are well-motivated DM candidates.
They are expected to either self-annihilate or decay into Standard Model particles, such as neutrinos, antiprotons, electrons/positrons, photons and so on. One kind of methods for the DM identification, namely DM indirect detection, is to search such signals from DM annihilation or decay. Of particular interest is the gamma-ray observation with high sensitivity. Since the propagation process is simple and the energy loss is small, the photons are very powerful probes to reveal the DM property.

Recently, the Fermi-LAT collaboration reported their 4-year measurement of the diffuse isotropic gamma-ray background (IGRB) at high latitudes with $|b| > 10^{\circ}$ \citep{2015ApJ...799...86A}. Compared with the previous measurements \citep{1998ApJ...494..523S,2010PhRvL.104j1101A}, the new Fermi-LAT data further extend to higher energy range, from $0.1$ GeV to $820$ GeV, nearly four decades. Especially above $300$ GeV, a significant high energy cut-off has been discovered. The whole spectrum can be well described by a single power-law plus an exponential cutoff with the index $\gamma \sim 2.32\pm 0.02$ and E$_{\rm cut} \sim 279\pm 52$ GeV. The dominant component of the IGRB is believed to be originated from the extragalactic astrophysical sources, most of which are too faint or too diffuse to be resolved, such as blazars, mis-aligned active galactic nuclei and star-forming galaxies and so on. Some galactic sources, such as millisecond pulsars, can also contribute to the IGRB \citep{2015PhR...598....1F}. However, since the predicted intensity from astrophysical sources is highly model dependent, there still exists a possible contribution from DM annihilation or decay in the IGRB. Thus the IGRB is often considered to be a powerful probe to search for the DM signals, and has been used to set upper limits on the DM annihilation cross section or decay lifetime in several studies \citep{2002PhRvD..66l3502U, 2003MNRAS.339..505T, 2009JCAP...07..020P, 2009PhRvD..80b3517K, 2010JCAP...04..014A, 2010JCAP...10..023Y, 2010NuPhB.840..284C, 2010JCAP...01..023C, 2012JCAP...11..021B, 2012PhRvD..86h3506C, 2012PhRvD..85d3509A, 2014PhRvD..89b3012B, 2014JCAP...02..014C, 2015arXiv150105464T, 2015ApJ...800L..27A, 2015arXiv150105316D, 2015arXiv150202007A, 2015ApJS..217...15X, 2015arXiv150601030C}.

On the other hand, a hot issue has received considerable attention in the DM study. In recent years, several experiments, such as PAMELA \citep{2009Natur.458..607A}, ATIC \citep{2008Natur.456..362C} and Fermi \citep{2009PhRvL.102r1101A}, reported an excess of the comic ray electron-positron measurement. Most recently, the AMS-02 results \citep{2014PhRvL.113l1101A, 2014PhRvL.113l1102A} have confirmed such excess from $\sim 0.5- 500$ GeV with a high precision. This anomaly can be explained by the DM with a large annihilation cross section to charged leptons, which is several orders of magnitudes over the thermal freeze-out value $3\times 10^{-26}$ cm$^{-3}$s $^{-1}$. Such DM particles would also inevitably induce significant gamma-ray signals by the cascade decay, internal bremsstrahlung, final state radiation (FSR), and the inverse Compton scattering (ICS) of electrons to background radiation field. Therefore, the IGRB is naturally summoned up as a powerful tool to constrain the DM explanations of the positron excess.

In this work, we study the constraints on the DM annihilation cross section and decay lifetime by using the latest Fermi-LAT IGRB results, and compare these limits with the DM parameter space which can explain the latest AMS-02 electron-positron observation. Compared with the previous works, we have made following improvements:
\begin{enumerate}
\item Both extragalactic and galactic contributions of DM annihilation/decay are reckoned. The steady-state spatial distribution of electrons and corresponding ICS gamma-rays in the Galaxy are computed by GALPROP \footnote{http://galprop.stanford.edu} \citep{1998ApJ...493..694M,1998ApJ...509..212S} with the comprehensive consideration of the transport equation and the background radiation field.
\item Three kinds of limits, namely conservative, background-fixed, and background-relaxed, are adopted and compared with each other. The goodness of bound depends on the limit method. Especially we show that the shape of bound curves could vary with constraint methods.
\item New cosmic-ray data have been extensively applied. We consider recent AMS-02 proton \citep{2014arXiv1402.0467C}, B/C \citep{ams02web} and electron-positron data \citep{2014PhRvL.113l1101A, 2014PhRvL.113l1102A}. They are used to constrain transport parameters in the Galaxy \citep{2015PhRvD..91f3508L, 2015arXiv150407230L}, and obtain the updated DM parameter space favored by cosmic-ray positron anomaly.
\end{enumerate}

The paper is organized as follows: In section II, we give a comprehensive introduction to the gamma-ray flux from DM annihilation(or decay). For the extragalactic DM annihilation, the dominant theoretical uncertainties arise from the unclear clustering history and properties of small DM halos. We consider these uncertainties under different assumptions about minimum DM halos. In section III, we discuss the limit approach to the DM annihilation(or decay). Then we illustrate our analysis of results. We derive the constraints under some different concentration models in DM annihilation. But for the decaying DM, due to that there are no above uncertainties, the constraints are quite confirmative. We also use GALPROP to calculate the propagation of the DM induced electrons and positrons, and obtain the parameter space accommodating the AMS-02 results. We compare these parameter space with the IGRB constraints. Finally, the summary are given in Section IV.

\section{Diffuse Gamma-Rays from Dark Matter Annihilation/Decay}
Both the extragalactic and galactic DM can produce high energy photons. The gamma-ray flux induced by extragalactic DM depends on the history of DM clustering and is essentially isotropic. On the other hand, the spatial distribution of the galactic gamma-ray signal is apparently anisotropic due to our special position in the Galaxy. Even after rigorously subtracting anisotropic component of galactic gamma-rays, there would still exist an residual isotropic component in the IGRB, which is equal to the signal from the direction of anti galactic center. Thus both the extragalactic and galactic DM would contribute to the IGRB signal, and the expected DM-induced IGRB flux can be written as \citep{2010NuPhB.840..284C, 2012PhRvD..86h3506C, 2012PhRvD..85d3509A}
\begin{equation}
\Phi^{\text{DM}} = \Phi^{\text{DM}}_{\text{EG}} +\Phi^{\text{DM}}_{\text{G}}{\Big|_{\text{antiG}}}.
\label{eqn:Phi_DM}
\end{equation}

\subsection{Gamma-Rays from Cosmological Dark Matter Evolution}
\label{DGBCDM}
The total gamma-ray flux emitted from the extragalactic annihilating DM at different redshifts is given by
\cite{2002PhRvD..66l3502U, 2010JCAP...10..023Y, 2012PhRvD..85d3509A},
\begin{equation}\label{Phi_EG}
\Phi^{\rm anni}_{\text{EG}}(E,z) = \frac{c(1+z)^2}{4\pi}
\frac{\Omega_\chi^2\rho_c^2\langle\sigma v\rangle}{2m_\chi^2}
\int_z^{\infty}{\rm
d}z'\frac{(1+z')^3[\Delta^2(z')+1]}{H(z')}\frac{{\rm d}N}{{\rm d}E'}
\exp\left[-\tau(z;z',E')\right],
\end{equation}
where $m_\chi$ is the mass of DM particle, and $\langle\sigma v\rangle$ is the corresponding thermal averaged annihilation cross section. $H(z)=H_0\sqrt{
(\Omega_\chi+\Omega_b)(1+z)^3+\Omega_\Lambda}$ and $\rho_c=3H_0^2/8\pi G$ are the Hubble parameter at redshift $z$ and current critical density of the Universe, respectively. For the latest cosmological parameters $\Omega_{\chi}$, $\Omega_{b}$, $\Omega_{\Lambda}$ and $h$, we refer to the values from \citep{2014ChPhC..38i0001O}. $\Delta^2(z)$ denotes the enhancement of DM annihilation, and will be introduced in a great detail in the next subsection. In eq. (\ref{Phi_EG}), $z$ and $z'$ are redshifts at which photons are observed and emitted respectively. ${\rm d}N/{\rm d}E'$ indicates the initial gamma-ray spectrum per DM pair annihilation, and $E'\equiv E(1+z')/(1+z)$ is the photon energy at redshift of emission $z'$. The prompt photons from DM annihilation are produced by the final-state radiations or cascade decays of the annihilation products. In this work, the injected energy spectrum of prompt photons is generated by PPPC4DMID\citep{2011JCAP...03..051C}.

The photons can also come from the ICS by DM-induced electrons and positrons off the interstellar radiation field, such as the cosmic microwave background(CMB), infrared photons and starlight. The gamma-ray flux from the ICS process is given by
\begin{equation}
\left.\frac{{\rm d}N}{{\rm d}E}\right|_{\rm IC}= c \int {\rm d}\epsilon\,
n(\epsilon)\int {\rm d}E_e \frac{{\rm d}n}{{\rm d}E_e} \times
F_{\rm KN}(\epsilon,E_e,E),
\end{equation}
where $n(\epsilon)$ is the number density distribution of the background radiation as a function of energy $\epsilon$ at redshift $z$. For the cosmological ICS process, we only take into account the CMB photons. $\dif n/\dif E_e$ is the energy spectrum of electrons. In this work, we adopt the assumption that electrons quickly lose their energy and the resulting distribution of electrons reaches equilibrium \citep{2009JCAP...07..020P, 2010JCAP...10..023Y}. Hence the spectrum is evaluated by equating the injected rate of DM electrons with the corresponding energy loss rate, which can be written as
\begin{equation}
\frac{{\rm d}n}{{\rm d}E_e}=\frac{1}{b(E_e,z)}\int_{E_e}^{m_\chi}
{\rm d}E_e'\frac{{\rm d}N_e}{{\rm d}E_e'} .
\end{equation}
with the energy loss rate $b(E_e,z)\approx 2.67\times 10^{-17}(1+z)^4
\left(E_e/{\rm GeV}\right)^2$ GeV s$^{-1}$. The differential Klein-Nishina cross section $F_{\rm KN}(\epsilon,E_e,E)$ is adopted as the following form \citep{1968PhRv..167.1159J, 1970RvMP...42..237B}
\begin{equation}
F_{\rm KN}(\epsilon,E_e,E)=\frac{3\sigma_T}{4\gamma^2\epsilon}\left[2q
\ln{q}+(1+2q)(1-q)+\frac{(\Gamma q)^2(1-q)}{2(1+\Gamma q)}\right],
\end{equation}
where $\sigma_T$ is the Thomson cross section, $\gamma$ is the Lorentz factor
of electron, $\Gamma=4\epsilon\gamma/m_e$, and $q=E/\Gamma(E_e-E)$. On a separate note, when $q<1/4\gamma^2$ or $q>1$, $F_{\rm KN}(\epsilon,E_e,E)=0$.

\subsection{Clumpiness Factor of Dark Matter Annihilation}
\label{CLUMPY}
As the DM annihilation rate is proportional to the square of number density, $\rho^2$, the annihilation signal would be significantly enhanced in the clumpy halos. The enhancement factor $\Delta^2(z)$ can be defined as summing up the contributions of all the halos with different masses formed in the history of Universe
\begin{equation}
\Delta^2(z)=\frac{\Delta_{\rm vir}(z)}{3\rho_\chi}\int{\rm d}M_{\rm vir}
M_{\rm vir}\frac{{\rm d}n(z)}{{\rm d}M_{\rm vir}} \frac{\int\tilde
{\rho}^2(x)x^2{\rm d}x}{\left(\int\tilde{\rho}(x)x^2{\rm d}x\right)^2}
x_{\rm max}^3 ~,
\end{equation}
where $M_{\rm vir}$ is the virial mass of the DM halo, ${\rm d}n(z)/{\rm d}M_{\rm vir}$ is the halo mass function, and $\tilde{\rho}$ is defined to describe the inner density profile of a single DM halo. Due to the self-similarity in the halo formation, all the halos share a common profile. Here we adopt the well-known NFW profile \citep{1996ApJ...462..563N}
\begin{equation}
\label{rho_tilde}
\tilde{\rho}(x) = \frac{\rho}{\rho_s} = \frac{1}{x (1+x)^2}
\end{equation}
with $x\equiv r/r_s$. The scale radius $r_s$ is related to the virial radius $r_{\rm vir}$ through
\begin{equation}
\label{r_s}
r_s = \frac{r_{\rm vir}}{c_{\rm vir}}.
\end{equation}
The virial radius $r_{\rm vir}$ can be directly derived from the virial mass $M_{\rm vir}$
\begin{equation}
r_{\rm vir}=\left(\frac{3M_{\rm vir}}{4\pi\Delta_{\rm vir}(z)\rho_\chi(z)}
\right)^{1/3},
\end{equation}
where $\rho_{\chi}(z) = \rho_{\chi}(1+z)^3$ is the mean DM density at redshift $z$.
The virial overdensity $\Delta_{\rm vir}(z)$ is taken to be \cite{1998ApJ...495...80B}
\begin{equation}
\Delta_{\rm vir}(z)=(18\pi^2+82y-39y^2)/(1+y),
\end{equation}
with $y=\Omega_m(z)-1$ and $\Omega_m(z)=\Omega_m(1+z)^3/(\Omega_m(1+z)^3+
\Omega_\Lambda)$.

In eq. (\ref{r_s}), the concentration parameter $c_{\rm vir}$ is a function of the virial mass $M_{\rm vir}$ and redshift $z$. The value of $c_{\rm vir}$ is usually obtained from the N-body simulation. However, the halos with low masses are beyond the power of the state-of-the-art resolution. Thus their $c_{\rm vir}$ is roughly evaluated by the extrapolation according to the fitting formula within the reach of simulations. The DM-induced gamma-ray flux can be enhanced by the promoted concentration as a result of the larger annihilation rate. In the cold dark matter(CDM) scenario, the structures are organized by 'bottom-up' fashion, i.e. the smaller structures formed earlier than larger ones. Since those massive halos assemble later and experience recent major merger, they typically hold lower concentrations compared with those growing quiescently and with smaller mass. This means that the concentration varies inversely with the halo mass \citep{2010gfe..book.....M}. Thus the gamma-ray intensity is sensitive to the lower halo mass cut-off and the slope of the concentration model. Here we consider two concentration models: one is an analytical model developed in \citep{2001MNRAS.321..559B} (B01), and the other is a direct extrapolation of the fitting results from the simulation \citep{2008MNRAS.391.1940M}(M08). In the above models, we assume the linear redshift evolution of the concentration parameter, i.e. $c_{\rm vir}(z)=c_{\rm vir}(z=0)/(1+z)$
\cite{2001MNRAS.321..559B}.

\begin{figure}[!htb]
\includegraphics[height=5.cm,angle=0]
{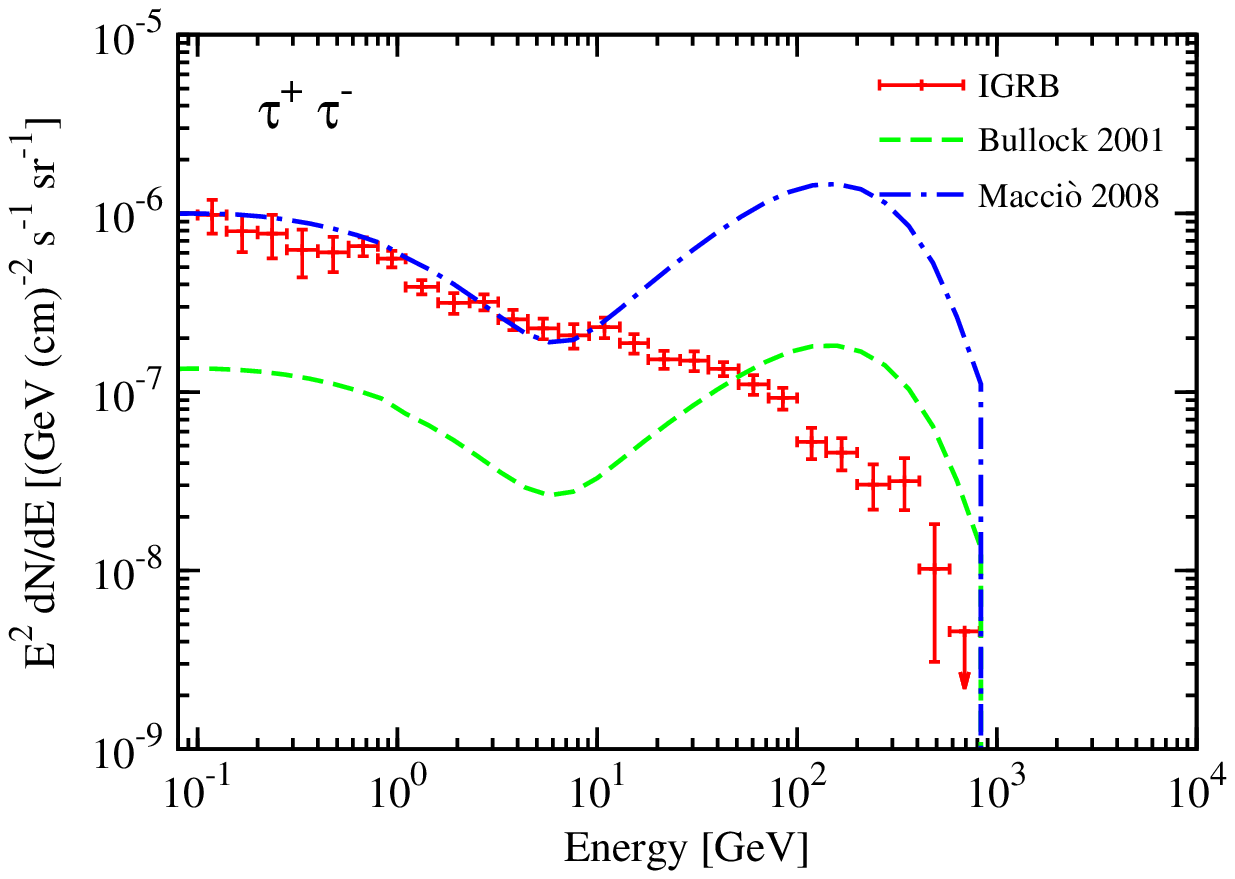}
\includegraphics[height=5.cm,angle=0]
{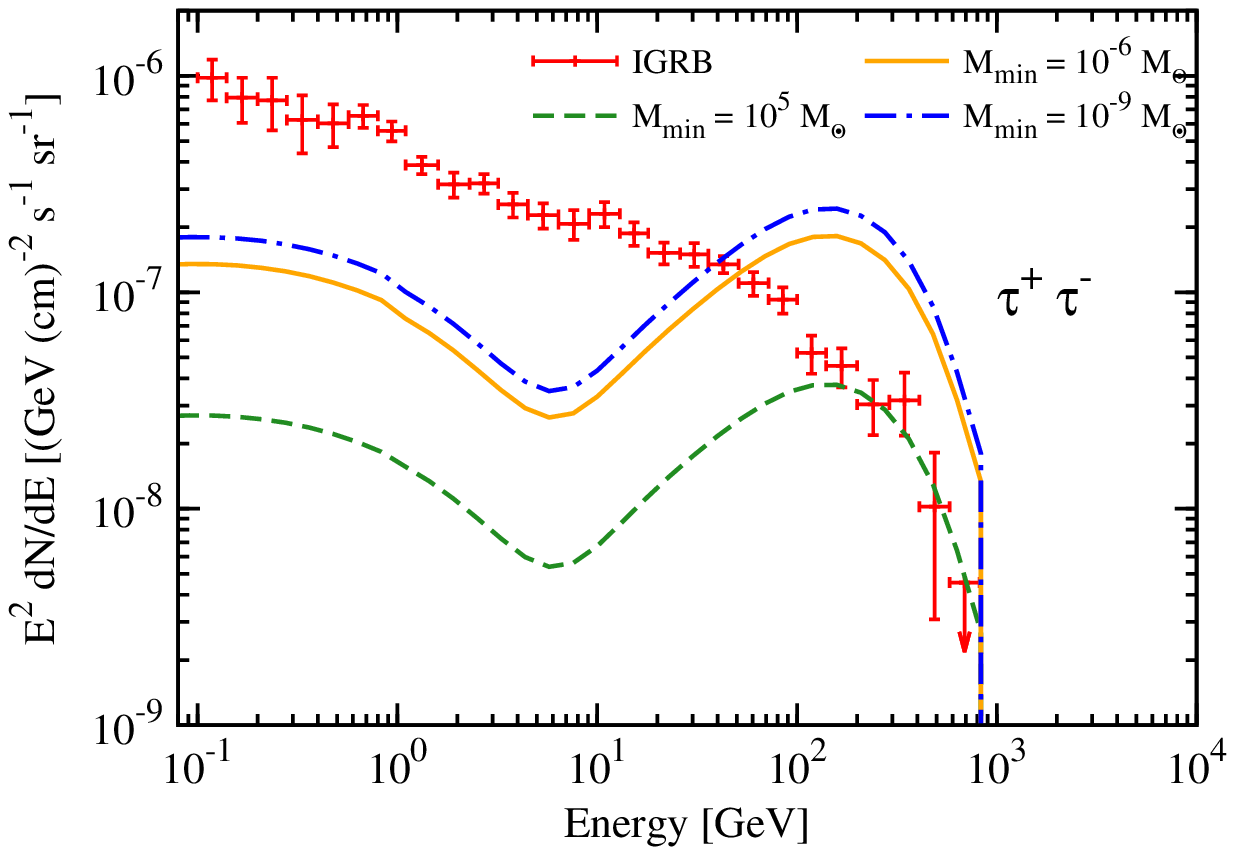}
\caption{
\label{fig:cmp_Cvir}
Left: the extragalactic gamma-ray spectra due to different sets of the concentration parameter $c_{\rm vir}$. The annihilation channel is chosen to be $\chi\chi \rightarrow\tau^+\tau^-$. The green dashed and blue dash-dot lines represent the spectra in the models of B01 \citep{2001MNRAS.321..559B} and M08 \citep{2008MNRAS.391.1940M}, respectively. Here the minimum DM halo mass is $M_{\rm min} = 10^{-6}$ M$_{\odot}$. Right: the same gamma-ray spectra assuming different minimum DM halo masses. The blue dash-dot, yellow solid and green dash lines correspond to $M_{\rm min} = 10^{-9}$, $10^{-6}$, $10^{5}$ M$_{\odot}$ in the B01 model.
}
\end{figure}

In this section, we compute the diffuse gamma-ray contributions of four DM benchmark points listed in Table. \ref{tab:bestfit_AMS02}. These parameter points are derived from a Markov Chain Monte Carlo (MCMC) fitting \citep{2010PhRvD..82b3506Y} to the latest AMS-02 electron/positron measurements \citep{2014PhRvL.113l1101A, 2014PhRvL.113l1102A}. In the left panel of Figure \ref{fig:cmp_Cvir}, we show the extragalactic DM-induced gamma-ray spectra under two different concentration models. The dashed and dash-dot lines represent the spectra in the models of B01 and M08, respectively. Here the minimum DM halo mass is taken to be $M_{\rm min} = 10^{-6}$ M$_{\odot}$. Although both B01 and M08 models provide a good fitting to the concentration parameters within the resolution of the N-body simulation, the different extrapolations in the low halo mass region still produce nearly one order of magnitude difference. On the other hand, the low mass cutoff of the DM halo is also unclear due to the limited resolution of the N-body simulation. In the right panel of Figure \ref{fig:cmp_Cvir}, we show the gamma-ray spectra for different assumptions of the minimum DM halo mass $M_{\rm min} = 10^{-9}$, $10^{-6}$, and $10^{5}$ M$_{\odot}$ in the B01 model. We can see that the gamma-ray intensity gradually raises with $M_{\rm min}$ decreasing. In the rest of this paper, we always take the minimum halo mass to be $M_{\rm min} = 10^{-6}$ M$_{\odot}$. Note that here the effect of extragalactic background light has been include, which will be expatiated in the next subsection.

\begin{table}[htbb]
\centering
\begin{tabular}{c | c c | c c}
\cline{1-5}
\multicolumn{1}{c| }{}
 & \multicolumn{2}{ c| }{Annihilation}
 & \multicolumn{2}{c}{Decay} \\
\cline{1-5}
\multirow{2}{*}{Channel} & $m_{\chi}$ & $\langle\sigma v\rangle$ & $m_{\chi}$ & $\tau$ \\
          & (GeV) & ($10^{-23}$ cm$^3$s$^{-1}$) & (GeV) & ($10^{26}$ s) \\
\hline
$\mu^{+}\mu^{-}$   & 417.44  & 0.30  & 808.63 & 9.13 \\
\hline
$\tau^{+}\tau^{-}$ & 1007.84 & 2.11 & 1774.76 & 3.21 \\
\hline
\end{tabular}
\caption{
\label{tab:bestfit_AMS02}
The best-fit values of mass-cross section(decay lifetime) parameter space for the latest AMS-02 positron-electron data \citep{2014PhRvL.113l1101A, 2014PhRvL.113l1102A}. The DM are chosen to annihilate(decay) into $\mu^{+}\mu^{-}$ and $\tau^{+}\tau^{-}$ channels.
}
\end{table}

\subsection{Extragalactic background light}
\label{EBL}
The factor $\exp[-\tau(z; z', E')]$ characterizes the absorption of gamma-ray photons when crossing the universe. $\tau(z; z', E')$ is the optical depth of gamma photons between observed redshift $z$ and emission redshift $z'$, and is obtained by following relation:
\begin{equation}
\tau(z;z',E')=c\int_z^{z'}{\rm
d}z''\frac{\alpha(E'',z'')}{H(z'')(1+z'')},
\end{equation}
where $E''=E'(1+z'')/(1+z')$, and $\alpha(E,z)$ is the absorption coefficient. As far as we are concerned, the dominant energy loss of high energy photons is the scattering with extragalactic UV background light. In this work, we refer to the UV background model given by \citep{2012MNRAS.422.3189G}. The UV background mainly affects the gamma-ray flux above $100$ GeV, which is suppressed by roughly one order of magnitude. Besides we still consider other energy loss processes: pair production on neutral matter($6 < z < 1000$), pair production on fully ionized matter($z < 6$), photon-photon scattering and photon-photon pair production with the CMB photons \citep{2011JCAP...03..051C}. These interactions give a very small contribution to the attenuation of high energy gamma-ray photons.

\begin{figure}[!htp]
\includegraphics[height=5.cm,angle=0]
{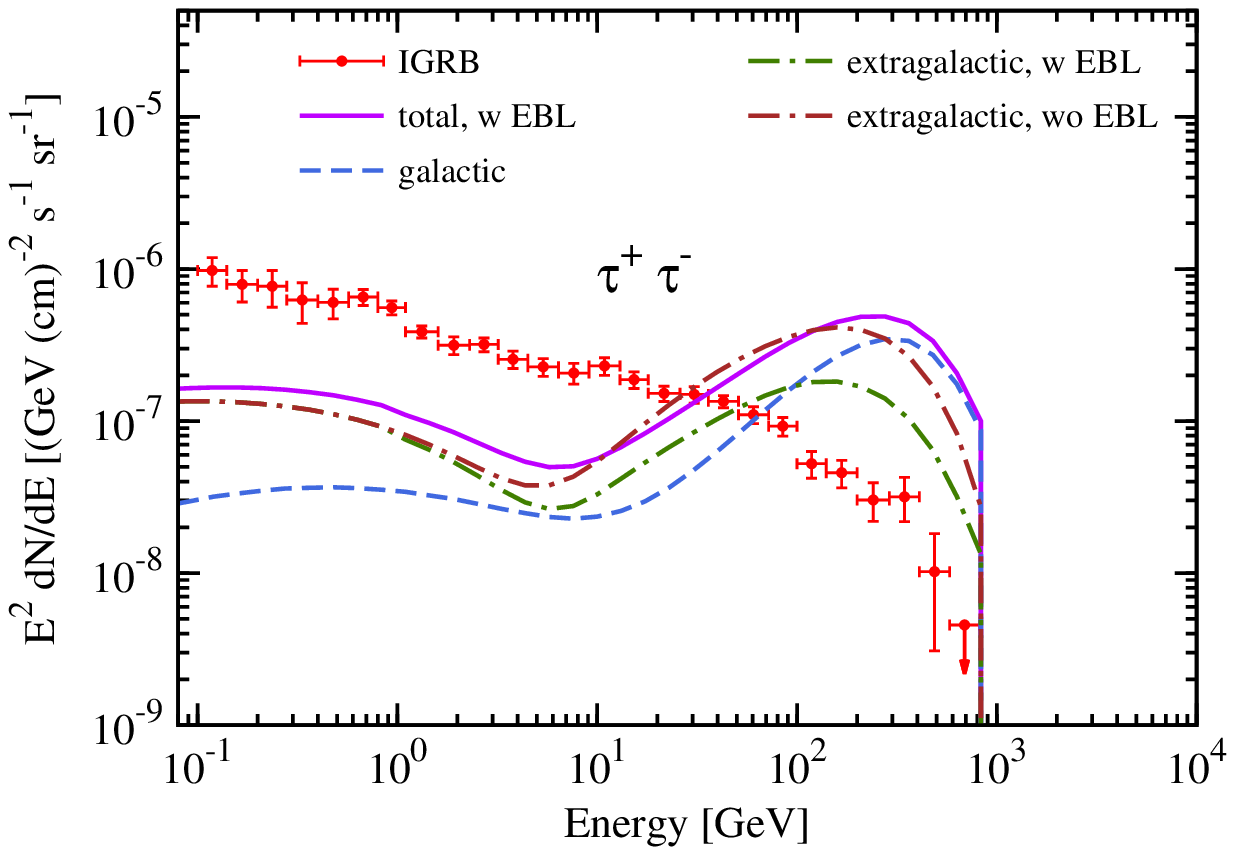}
\includegraphics[height=5.cm,angle=0]
{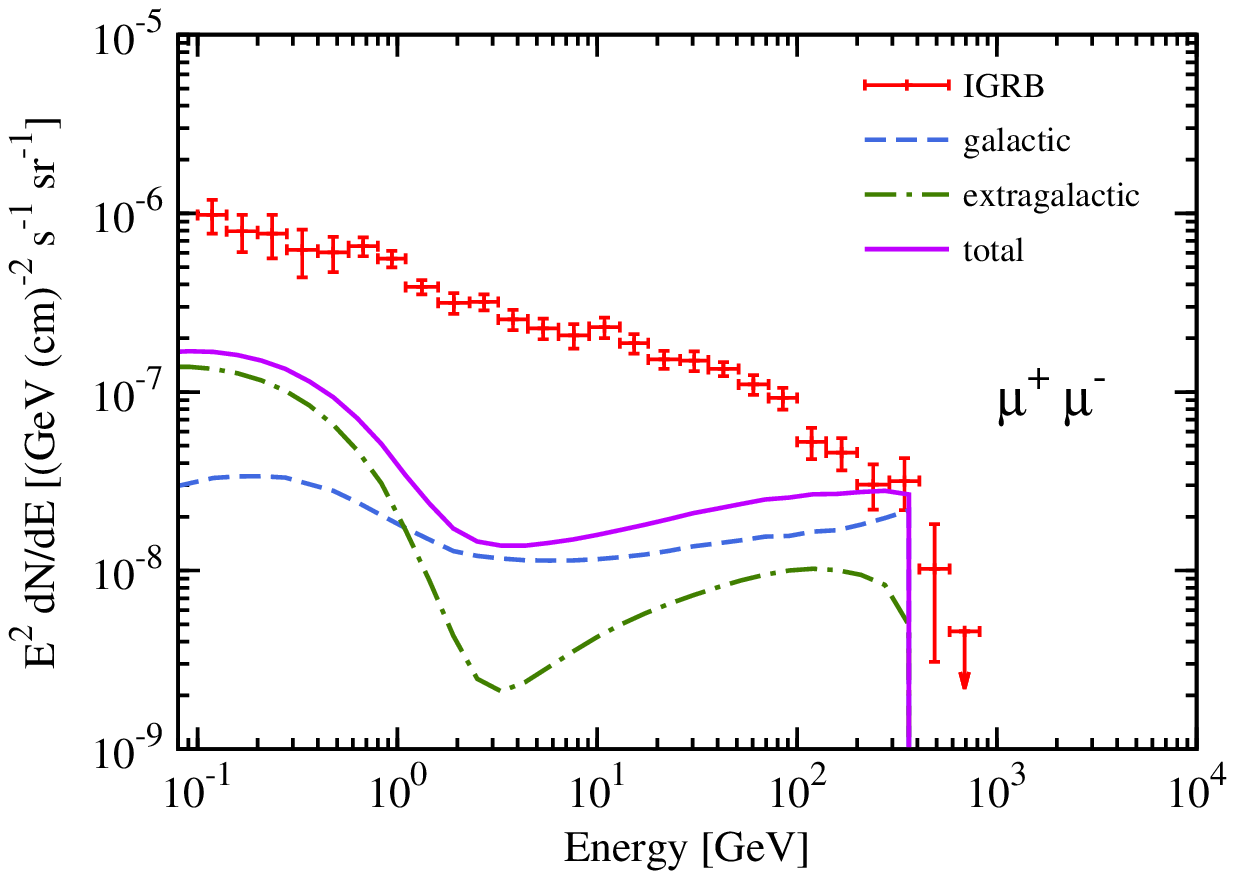}
\caption{
\label{fig:cmp_EBLsuppress}
Left: the influence of extragalactic background light (EBL). The green dash-dot and brown dash-dot lines represent the extragalactic flux with and without EBL. The blue dash line is the galactic contribution. The purple solid line is the total flux with EBL. The annihilation channel is $\chi\chi \rightarrow\tau^+\tau^-$. Right: the galactic(blue dash line), extragalactic(green dash-dot line) and total(purple solid line) gamma-ray flux from $\chi\chi \rightarrow \mu^+\mu^-$ channels of DM annihilation. Both DM particle's cross section and mass are listed in the table \ref{tab:bestfit_AMS02}.
}
\end{figure}

In the left panel of Figure \ref{fig:cmp_EBLsuppress}, we show the extragalactic gamma-ray spectra with and without the absorption effect of extragalactic background light. The annihilation is chosen to be $\chi\chi \rightarrow \tau^+\tau^-$ channel. Here the concentration model is chosen to be B01. It is apparent that EBL mainly influence the high energy gamma-ray spectra, above tens of GeV. The blue dash and purple solid lines are the galactic gamma-ray flux and total flux with EBL.

\subsection{Diffuse Gamma-Rays from Galactic Dark Matter annihilation}
\label{DGRGDM}
The gamma-ray signal from the annihilation of Galactic DM particles is obtained by the light-of-sight integral of squared DM density at an angle $\psi$ with respect to the direction of galactic center. The prompt radiation is given by
\begin{equation}
\label{fsrg}
\Phi^{\rm Prompt}_{\rm G}(E, \psi) = \langle\sigma v\rangle \frac{R_{\odot} \rho^2_{\odot}}{8\pi m_\chi^2}\frac{dN}{dE}  \int_{l.o.s.}{ \left[ \frac{\rho(r(x, \psi =\psi(b, \ell)))}{\rho_{\odot}} \right]^2 ~ \frac{\dif x}{R_{\odot}}} ~.
\end{equation}
$r(b, \ell,x) = \sqrt{R_\odot^2-2x R_\odot\cos(\ell)\cos(b)+x^2}$ is the distance to the galactic center, where $(b,l)$ are the galactic coordinates. Due to the finite resolution of the telescope, the gamma-rays actually are received from a finite observational solid angle. Therefore the predicated gamma-ray flux from DM annihilation should be averaged within a solid angle $\Delta \Omega$ toward an observational region
\begin{equation}
\label{FSR_G}
\bar{\Phi}^{\rm Prompt}_{\rm G}(E, \psi) = \langle\sigma v\rangle \frac{R_{\odot} \rho^2_{\odot}}{8\pi m_\chi^2}\frac{dN}{dE} \int_{\Delta \Omega} \frac{\dif \Omega}{\Delta \Omega}\int_{l.o.s.}{ \left[\frac{\rho(r(b, \ell, x))}{\rho_{\odot}}\right]^2~ \frac{\dif x}{R_{\odot}}} ~.
\end{equation}
We find that for the case of anti-galactic direction, this average brings about negligible improvement. For the density distribution of the Galactic DM halo, we still adopt the NFW density profile with fixing the local DM density $\rho(r = r_{\odot}) = 0.3 $ GeV/cm$^3$ and total DM mass within $60$ kpc $M(\leq 60 ~\rm kpc) = 4.7\times 10^{11}$ M$_{\odot}$, which means $r_s = 24.42$ kpc and $\rho_s = 0.184$ GeV/cm$^3$ \citep{2011JCAP...03..051C}.

For the gamma-rays from the ICS by DM-induced high energy electrons, we need to solve the transport equation of electrons in the Galaxy. However, the high energy electrons can only transport a few hundreds of parsecs due to the significant energy loss. Thus the observed electrons are mainly originated from the nearby sources. Unlike the extragalactic ICS process, the background photons include two additional components as well as the CMB photons: infra-red light from the absorption and re-emission of starlight by galactic dust and starlight from stars in the galactic disk. Both of them mostly distribute in the Galactic disk and are spatial dependent. Yet the usual analytical solutions of transport equation often make simplified assumption on the radiation field. In this work, the package GALPROP is used to numerically solve the transport and ICS processes of electrons, in which the spatial distribution of background radiations has been included. The spectra of initial electrons injected by DM are still evaluated by PPPC4DMID \citep{2011JCAP...03..051C}. The transport parameters are consistent with those used to explain the latest AMS-02 results \citep{2015PhRvD..91f3508L}.

For the galactic DM annihilation, we still consider the boost factor due to DM substructures. Many analytic arguments and numerical simulations have confirmed the presence of substructure in the galactic DM halo \citep{1997ApJ...490..493N, 2010MNRAS.404..502G, 2011MNRAS.415.2475M}. We refer to the analytic substructure model developed by \citep{2008PhRvD..77j3509K, 2010PhRvD..81d3532K}. This method can extend to the mass scales which are too small to be resolved by the numerical simulations.

In Figure \ref{fig:cmp_EBLsuppress}, we compare the galactic gamma-ray flux with the extragalactic contribution. The annihilation channels are respectively $\chi\chi \rightarrow\tau^+\tau^-$ and $\chi\chi \rightarrow\mu^+\mu^-$. For $\tau^+\tau^-$ channel, the prompt radiation makes the stronger contribution. The galactic contribution exceeds the extragalactic one at higher energy, about hundreds of GeV. But for $\mu^+\mu^-$ channel, the prompt radiation flux is significantly weaker than ICS flux, thus the galactic flux holds a dominant position from lower energy, about several GeVs.

\subsection{Gamma-rays from DM decay}
Compared with the annihilating DM, the gamma-ray intensity from the decaying DM is only proportional to the cosmological DM density $\rho_{\chi}$. Thus it does not suffer from enormous uncertainties, such as density profile of DM halo, the history of structure formation, concentration parameter, halo mass function and so on. In this case, the resulting predictions would be relatively more solid. The accumulated DM-induced gamma-ray flux during the evolution of universe is given by \citep{2010NuPhB.840..284C, 2010JCAP...01..023C, 2012PhRvD..86h3506C}
\begin{equation}
\Phi^{\rm dec}_{\rm EG} = \frac{c}{4\pi} \frac{\Omega_{\chi} \rho_c}{m_{\chi} \tau_{\rm dec}} \int \frac{\dif z'}{H(z')} \frac{{\rm d}N}{{\rm d}E'} \exp[-\tau(z; z', E')],
\end{equation}
with $\tau_{\rm dec}$ the decay lifetime of DM particle. For the prompt contribution from galactic DM decay, we just need to make following substitution in eq. (\ref{FSR_G}):
\begin{equation}
\frac{\rho^2 \langle\sigma v\rangle}{2m^2_{\chi}} \rightarrow \frac{\rho}{m_{\chi} \tau}~.
\end{equation}
The spatial distribution and energy spectrum of electrons from galactic DM decay, and the ICS contribution to photons are also evaluated by GALPROP.

\begin{figure}[!htp]
\includegraphics[height=5.cm,angle=0]
{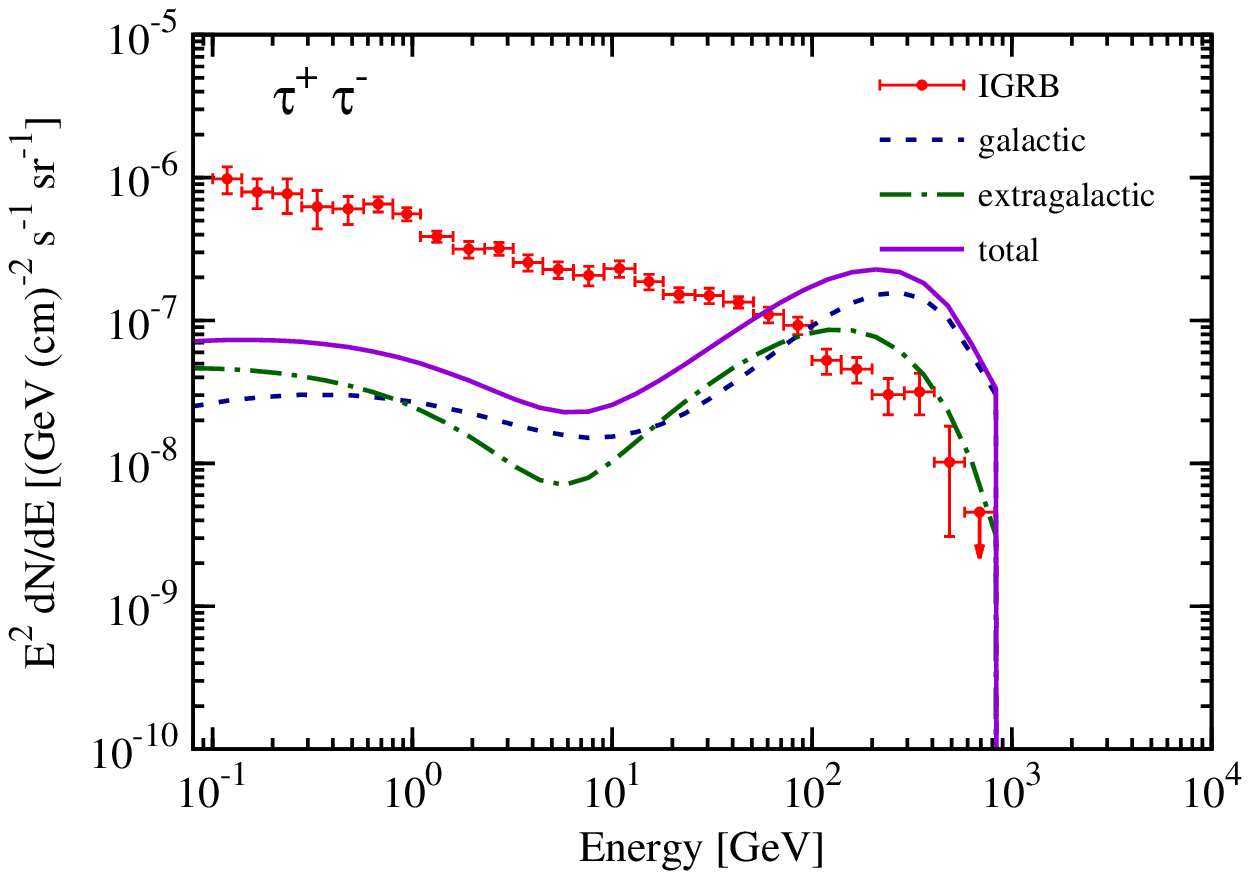}
\includegraphics[height=5.cm,angle=0]
{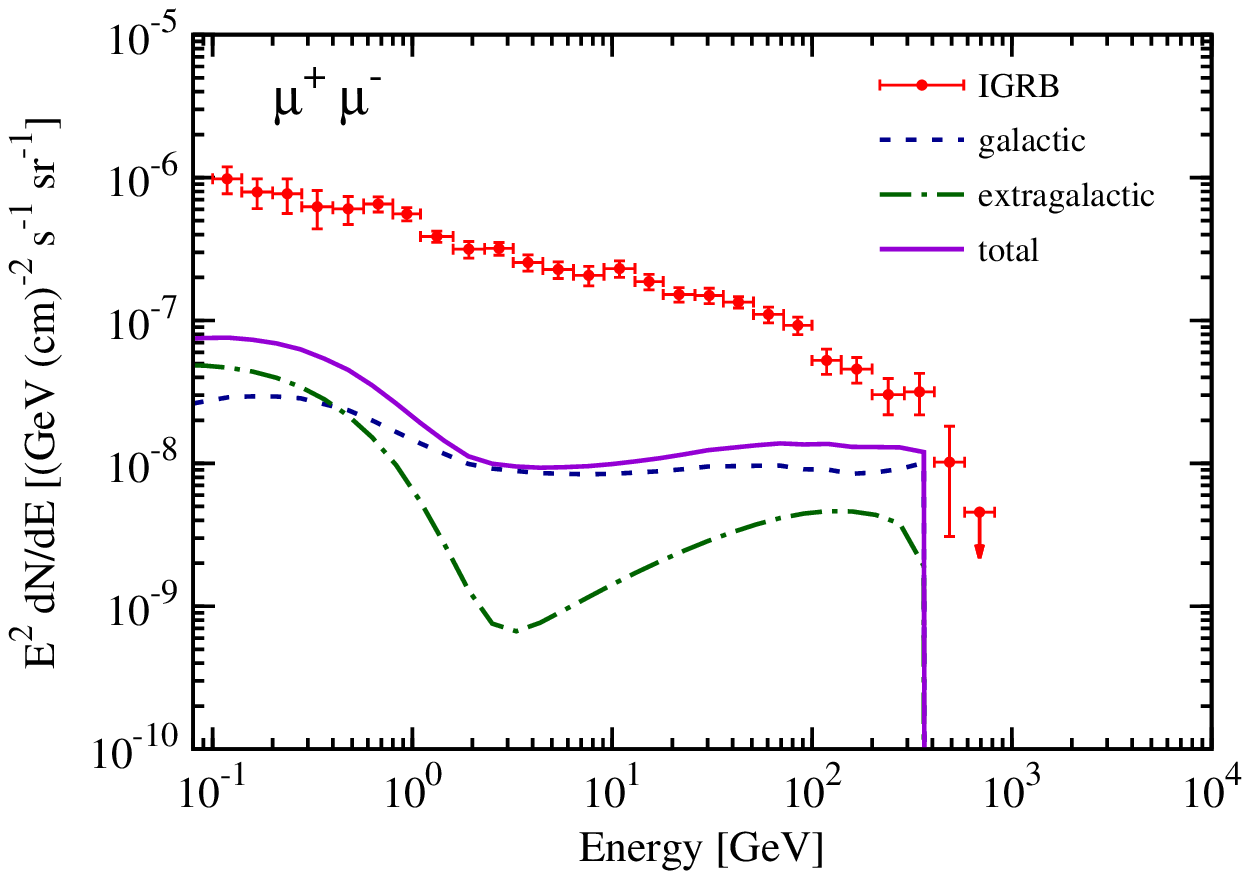}
\caption{
\label{fig:DGRS_decay}
The figure shows the galactic(blue dash line), extragalactic(green dash-dot line) and total(purple solid line) gamma-ray flux from different DM decay channels. The left is $\chi\chi \rightarrow\tau^+\tau^-$ channel, and the right is $\chi\chi \rightarrow\mu^+\mu^-$ channel. The cross section and mass are listed in the table \ref{tab:bestfit_AMS02}.
}
\end{figure}

The left and right panels of Figure \ref{fig:DGRS_decay} show the gamma-ray spectra for $\chi\chi \rightarrow\tau^+\tau^-$ and $\chi\chi \rightarrow\mu^+\mu^-$ channels, respectively. Both the galactic (short dash) and extragalactic(dash-dot) contributions are also shown.

\section{Constraints on Dark Matter Annihilation/Decay}

\subsection{Methods}
The main component of the observed IGRB is believed to be originated from unresolved astrophysical sources. In principle, the DM-induced signals can be obtained by subtracting all the astrophysical contributions from the Fermi-LAT data. The possible dominant candidates include blazars (including Flat Spectrum Radio Quasars and BL Lacertae) \citep{2011ApJ...737...80V, 2012ApJ...749..151Z, 2013MNRAS.431..997Z}, star-forming galaxies \citep{2012ApJ...755..164A, 2014ApJ...786...40L}, misaligned AGN \citep{2011ApJ...733...66I, 2014ApJ...780..161D}. Recent years, some authors have performed analysis by fitting the IGRB data with the astrophysical contributions along with their predicted theoretical uncertainties, and then set upper limits on the DM contribution\citep{2014PhRvD..89b3012B, 2014JCAP...02..014C, 2015ApJ...800L..27A, 2015arXiv150105316D, 2015arXiv150202007A}. Some studies claimed that the extragalactic gamma-ray background above $50$ GeV can be principally attributed to blazars \citep{2015arXiv151100693T}. However, the precise contributions of different populations are model dependent and remain unclear. In this work, we do not focus on the predictions and uncertainties of signatures from astrophysical sources, while adopting some model-independent methods to set constraints on DM annihilation/decay.

\textbf{Conservative limits:} As a first analysis, we require that the DM
contributions alone should not exceed the observed IGRB
spectra. The derived constraint is usually regarded
as the most conservative one.
The $\chi^2$ can be defined
in energy bins where the DM signal exceeds the IGRB intensity
\begin{equation}
\chi^2_{\text{cons}} = \sum_{i \in \{i|\phi^{\text{DM}}_i > D^{\text{max}}_i\}} \frac{[D^{\text{max}}_i -\phi^{\text{DM}}_{i}]^2}{\sigma^2_i}.
\end{equation}
$\phi^{\rm DM}_i$ is the DM-induced gamma-ray flux in the $i-$th energy
bin as a function
of $\left\langle \sigma v \right\rangle$ or $t_{\rm dec}$. We adopt the IGRB background based on the Galactic emission model A in Ref. \citep{2015ApJ...799...86A}.
We also incorporate foreground uncertainties into the IGRB spectra while
with unchanged $\sigma_i$ as Ref. \citep{2015arXiv150105464T}.
All these new data points are called $D^{\rm max}_i$. The corresponding $3\sigma$
DM limits are achieved when $\chi^2_{\rm cons} = 9$.

\textbf{Background fixed:} We assume a universal function to represent the total energy spectra from astrophysical sources. Its form is taken as a single power-law with an exponential cutoff at high energy
\begin{equation}
\phi^{\text{bg}} = I_0 \left(\frac{E}{100~\text{MeV}}\right)^{\gamma} \exp \left(-\frac{E}{E_c} \right),
\end{equation}
where $I_0$, $\gamma$, and $E_c$ are kept to be the best-fit values to the IGRB spectra under foreground model A \citep{2015ApJ...799...86A}. The DM-induced photon flux is assumed to be superimposed on the background flux. This method is widely employed in the past studies \citep{2010JCAP...04..014A, 2012PhRvD..86h3506C, 2015arXiv150105464T}. The $\chi^2$ is evaluated over all the energy bins:
\begin{equation}
\chi^2_{\text{sens}} = \sum_i \frac{[D_i -\phi^{\text{bg}}_i(I_0, \gamma, E_c) -\phi^{\text{DM}}_i]^2}{\sigma^2_i} .
\end{equation}
The $3\sigma$ limits are reached when
the DM signal component forces the $\chi^2$ to raise by more than $9$ with respect to the best-fit $\chi^2$ without DM signal.

\textbf{Background relaxed:} In this case, the astrophysical background is also assumed to be a single power-law plus an exponential cutoff, whereas $I_0$, $\gamma$, and $E_c$ are treated as free parameters as well as $m_{\chi}$ and $\left\langle \sigma v \right\rangle$(or $t_{\rm dec}$). For given $m_{\chi}$ and $\left\langle \sigma v \right\rangle$(or $t_{\rm dec}$), we can obtain a minimal $\chi^2$ via a global fitting to the IGRB data. The upperlimit on $\left\langle \sigma v \right\rangle$ (or $t_{\rm dec}$) can be obtained when corresponding $\chi^2$ deviates from the minimal value $\chi^2_{\rm min}$ by a particular value. Here GNU Scientific Library(GSL)\footnote{http://www.gnu.org/software/gsl/} is used to perform the nonlinear least-square fit.

\subsection{Results}

\begin{figure}[tbp]
\begin{center}
\includegraphics[height=19.cm,angle=0]
{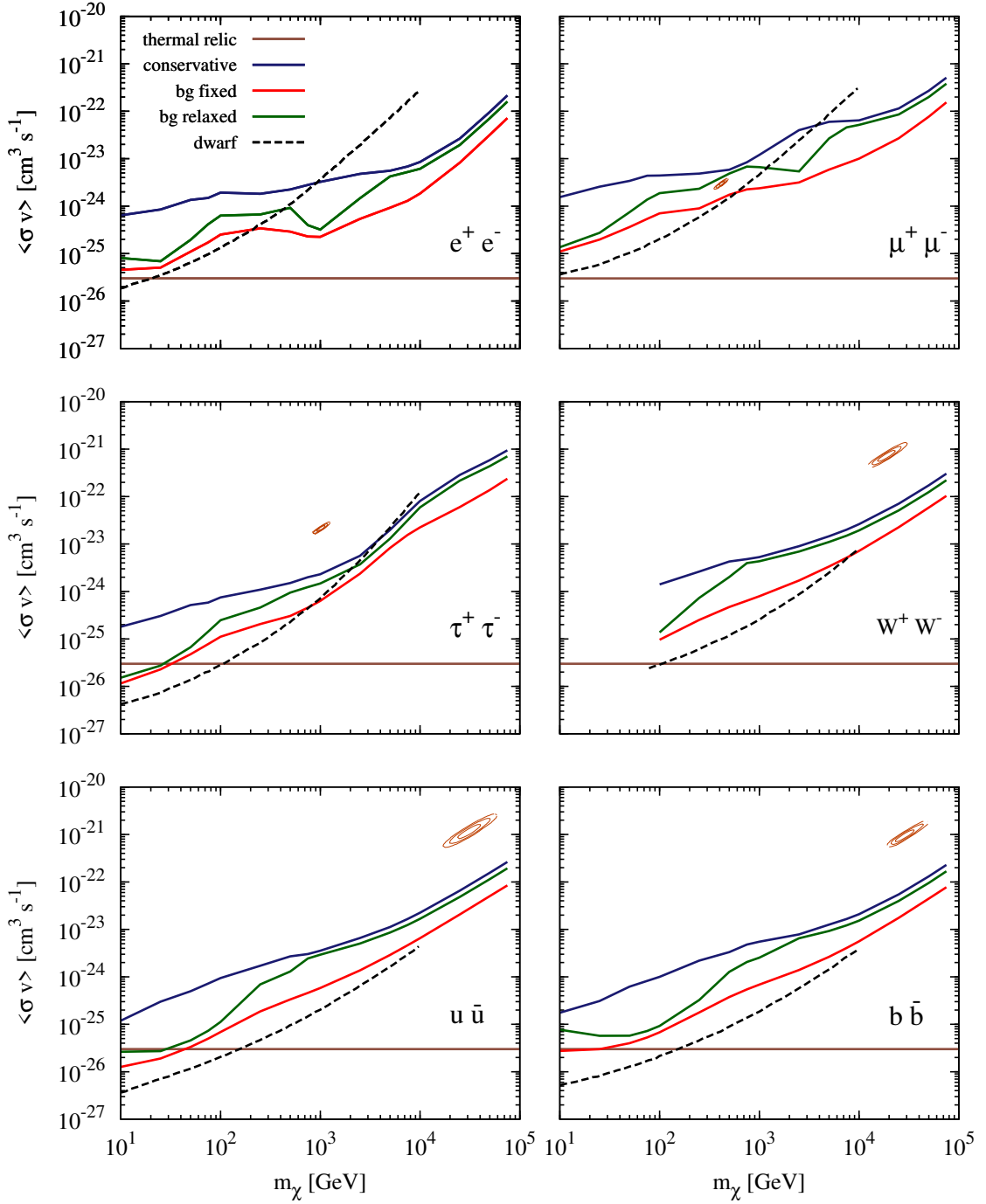}
\caption{
\label{fig:anni_B01}
The constraints on the DM annihilation cross section for six different DM annihilation channels: $e^+e^-$, $\mu^+\mu^-$, $\tau^+\tau^-$, $W^+ W^-$, $u\bar{u}$, and $b\bar{b}$.
The concentration model B01 \citep{2001MNRAS.321..559B} is adopted and $M_{\rm min} = 10^{-6}$ M$_{\odot}$. The blue, red and green solid lines denote the conservative, background-fixed and background-relaxed limits, respectively. The brown solid line denotes the nature annihilation cross section for the thermal relic density $\sim 3\times 10^{-26}$cm$^3$s$^{-1}$. Black dash lines are the constraints from the Fermi-LAT observations of dwarf spheroidal galaxies \citep{2014PhRvD..89d2001A}. The dark orange contours correspond to the $1\sigma$, $2\sigma$ and $3\sigma$ parameter regions accounting for the electron-positron excess observed by the AMS-02 \citep{2014PhRvL.113l1101A, 2014PhRvL.113l1102A}.
}
\end{center}
\end{figure}

In this section, we show the IGRB limits on the DM annihilation cross sections for six different channels: $e^+e^-$, $\mu^+\mu^-$, $\tau^+\tau^-$, $W^+ W^-$, $u\bar{u}$, and $b\bar{b}$ in Figure \ref{fig:anni_B01}.
Here we adopt the concentration model B01 \citep{2001MNRAS.321..559B} and set M$_{\rm min}$ to be $10^{-6}$ M$_{\odot}$. Three types of the curves represent the constraints of conservative(blue), background-fixed(red) and background-relaxed(green) methods, respectively. We can see that, compared with the conservative limits, the background-fixed limits on the DM annihilation cross section can be improved by about one order of magnitude in the mass region of $\sim \mathcal{O}(10^2)$~GeV.
The background-relaxed limits are always sandwiched between the conservative and background-fixed limits. For low DM masses, they are as stringent as the background-fixed limits. For the $\tau^+\tau^-$ and $u\bar{u}$ channels, these limits could even reach the thermal cross section $\langle \sigma v\rangle \sim 3\times 10^{-26}$cm$^3$s$^{-1}$ at the mass region of $\sim \mathcal{O}(10)$~GeV. When the DM mass increases, all the constraints become loose and their distinctions decreases. As can be seen that the background-relaxed limits tend to the conservative limits at the DM mass region of $\mathcal{O}(10)$~TeV.

For comparison, the constraints from the latest Fermi-LAT observations of dwarf galaxies \citep{2014PhRvD..89d2001A} are also shown in Figure \ref{fig:anni_B01}. For the hadronic channels,
the IGRB limits are always weaker than those of dwarf galaxies at low DM mass region. However, the IGRB observations could set stringent bounds for heavy DM particles annihilating to leptons as a result of large contributions from ICS processes. This is particularly clear for the $e^+e^-$ channel as shown in Figure \ref{fig:anni_B01}.

In Figure \ref{fig:anni_B01}, we also show the parameter regions accounting for the cosmic-ray electron-positron anomaly renewed by the AMS-02 collaboration.
The favored DM annihilation cross section and DM mass are derived from a global MCMC fit to the AMS02 data. We use the GALPORP to deal with the transport effect, and adopt a conventional diffusion-convection model. More comprehensive discussions can be available in Ref. \citep{2015PhRvD..91f3508L}. Here we do not consider the $e^+e^-$ final states, since the corresponding sharp electron-positron spectra cannot fit the current AMS-02 data. As shown in figure \ref{fig:anni_B01}, the available regions for $\mu^+\mu^-$ and $\tau^+\tau^-$ channels are much smaller than those for hadronic channels;DM masses required by the leptonic channels are also smaller than those for hadronic channels. We can see that almost all the channels have been excluded by the background-fixed IGRB limits. Only the parameter region for the $\mu^+\mu^-$ channel remains valid by the conservative IGRB limit.

\begin{figure}[tbp]
\begin{center}
\includegraphics[height=19.cm,angle=0]
{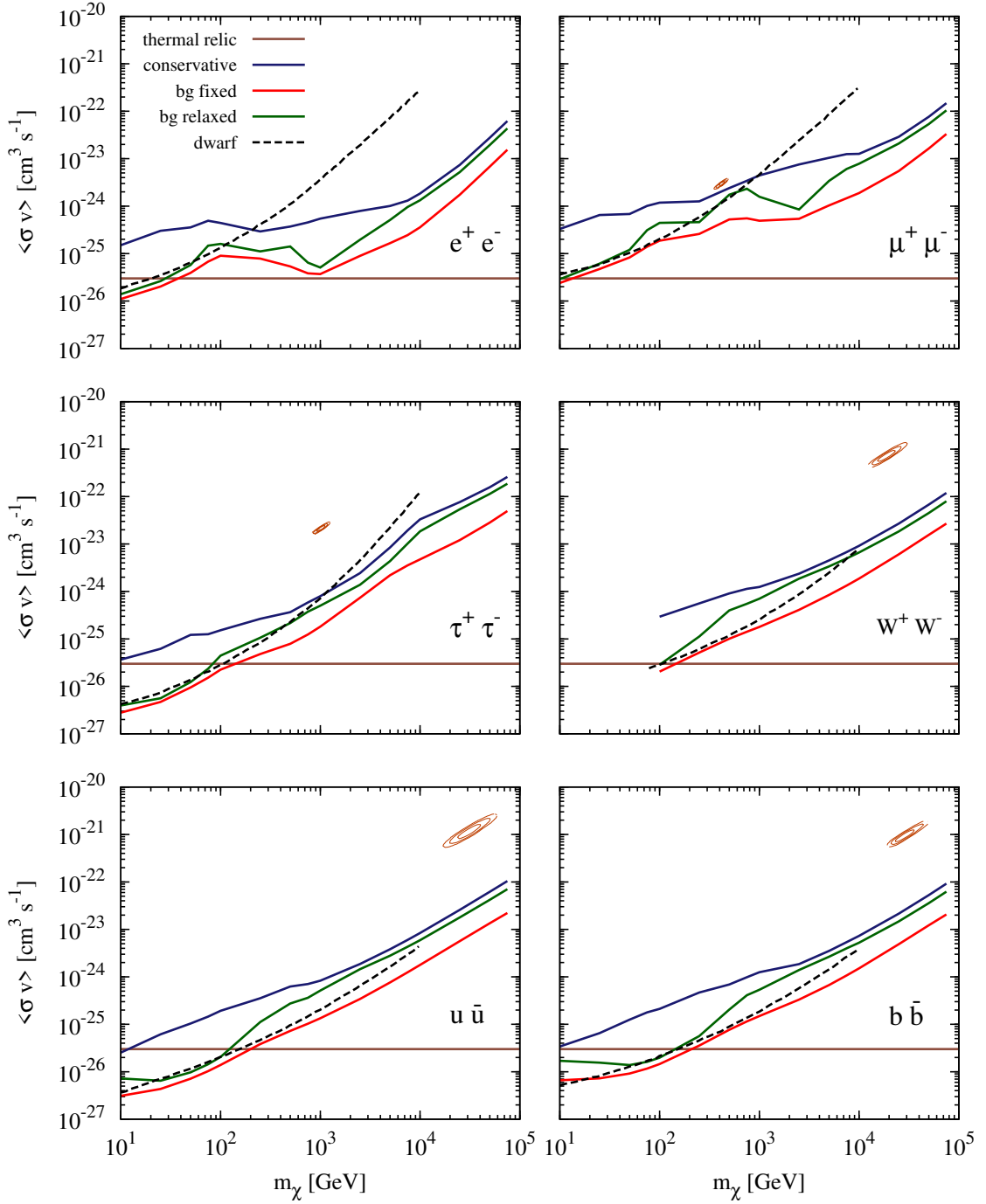}
\caption{
\label{fig:anni_M08}
The constraints on the DM annihilation cross section for six different DM annihilation channels. The concentration model M08 \citep{2008MNRAS.391.1940M} is adopted. The notations are the same as Fig.~\ref{fig:anni_B01}.
}
\end{center}
\end{figure}

In Figure \ref{fig:anni_M08} we show the IGRB limits on the DM annihilation cross section  for the concentration model M08 \citep{2008MNRAS.391.1940M}. All the limits are improved by almost one order of magnitude. This can be understood by the energy spectra shown in Figure \ref{fig:cmp_Cvir}. At low DM masses, the IGRB limits are already comparable to those from dwarf galaxies\cite{2014PhRvD..89d2001A}, which tend to $10^{27}$ cm$^3$s$^{-1}$. In this case, even the parameter space favored by the positron anomaly in $\mu^+\mu^-$ channel has been excluded readily by the conservative IGRB limit.

\begin{figure}[tbp]
\begin{center}
\includegraphics[height=19.cm,angle=0]
{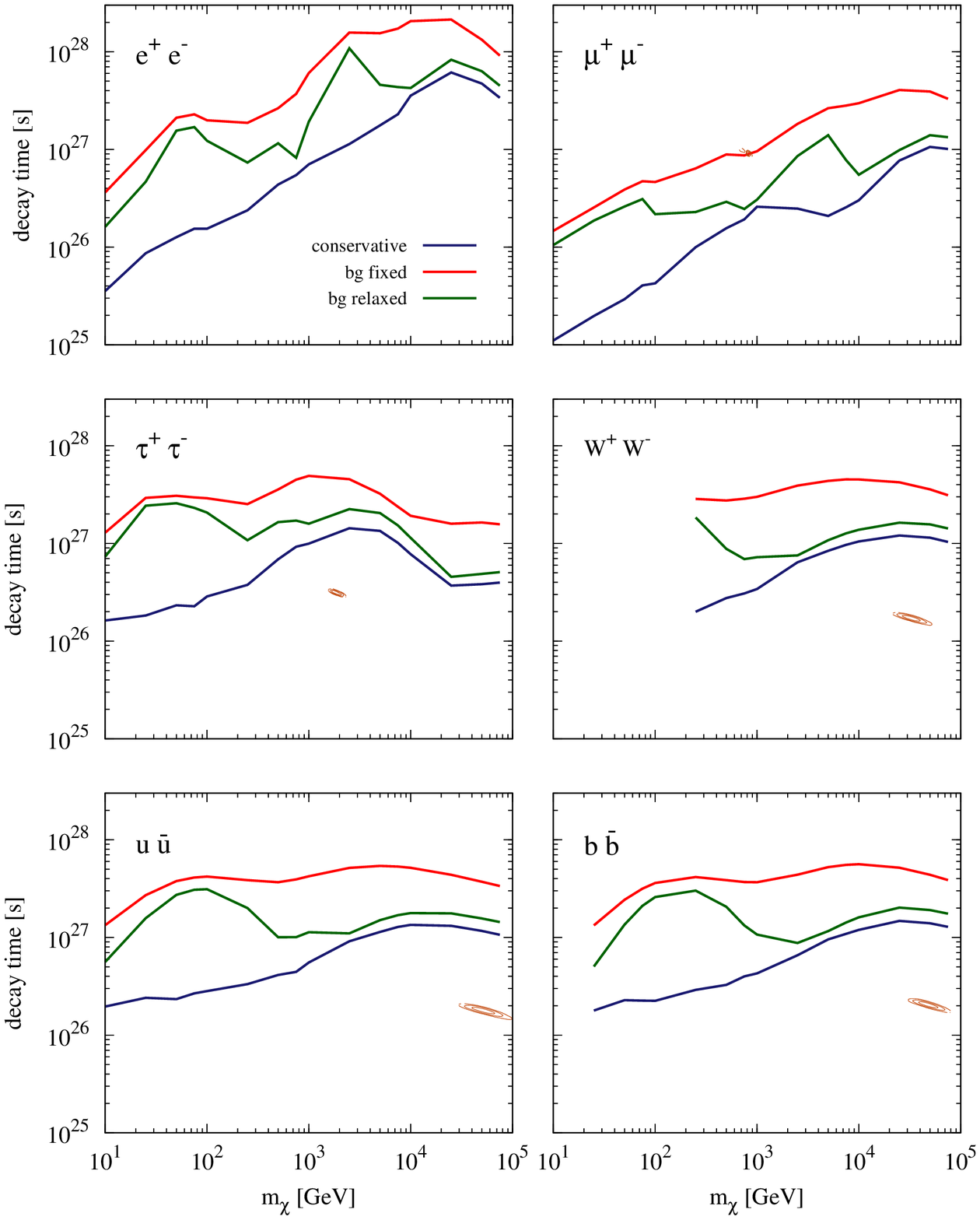}
\caption{
\label{fig:decay}
The constraints of dark matter decay channels. Blue solid line is conservative limit, red solid line is the limit of background-fixed and green solid line is background-relaxed limits. The dark orange contours are the parameter space favored by the cosmic-ray electron-positron excess.
}
\end{center}
\end{figure}

In Figure \ref{fig:decay}, we present the constraints on the lifetime of decaying DM. In contrast to DM annihilation, the gamma-ray fluxes generated by decaying DM are not significantly affected by the history of the structure formation. Therefore the constraints on the DM lifetime are more credible. The most stringent constraint comes from the $e^+e^-$ channel, and reaches $\tau \sim 10^{28}$s for the DM masses of $\mathcal{O}(1)$~TeV. This is because the main contributions of these two channels are photons from ICS and final state radiation processes, while the peak of energy spectra at high energies would become significant and easily constrained by Fermi-LAT data when DM mass increases. For the remaining channels, the limits are also stringent for low DM mass due to the contributions from cascade decay and hadronization processes. The regions in parameter space favored by the positron anomaly are also manifested. As can be seen that all the channels are disfavored by the background-fixed limits. Only the $\mu^+\mu^-$ channel remains allowed by the conservative limit.

\section{Summary}
\label{sum}
In this work, we use the latest Fermi-LAT IGRB data to set upper limits on the DM annihilation cross section or the DM lifetime for six channels, i.e. $e^+e^-$, $\mu^+\mu^-$, $\tau^+\tau^-$, $W^+ W^-$, $u\bar{u}$, and $b\bar{b}$. In order to consider the uncertainties from the multiplier of the extragalactic gamma-ray flux, the DM annihilation constraints are investigated in two competing parameterized concentration models, i.e. B01 \citep{2001MNRAS.321..559B} and M08 \citep{2008MNRAS.391.1940M}. In our analysis, we derive three kinds of limits, namely conservative, background-fixed and background-relaxed limits. Compared with the conservative method, the background-fixed method can improve the constraints by about one order of magnitude at low DM masses. If a combined fit accounting for both DM-induced flux and the astrophysical background is performed, the corresponding background-relaxed limits always lie between the conservative and background-fixed limits.

For DM annihilation, we find the most stringent bounds are for $\tau^+\tau^-$ and $u\bar{u}$ channels.
In the concentration parameter model is M08, the background-fixed limits for these two channels can be to the limits from the Fermi-LAT dwarf spheroidal galaxy observations in the mass region of $m_{\chi}\leq \mathcal{O}(10)$~GeV. For large DM masses $\sim \mathcal{O}(1)$TeV, the constraints for the leptonic channels can be stronger than the dwarf galaxy limits. This indicates that the IGRB is suitable to search for heavy DM.

We also investigate the IGRB constraints on the parameter regions favored by the cosmic-ray electron-positron excess. We find that almost all the annihilation channels
have been excluded by the background-fixed limits. Only the $\mu^+\mu^-$ channel remains valid by the conservative limit in the concentration model B01.
For decaying DM, the most stringent constraint is set for the $e^+e^-$ channel, which can reach even $\tau\sim 10^{28}$ s above several hundreds of GeV. Most decay channels favored by electron-positron anomaly have also been excluded by the conservative limits except the $\mu^+\mu^-$ channel. But the background-fixed limit is close to the border of its $3\sigma$ contours. The future observations will place more stringent constraint on this channel.

\section*{Acknowledgement}
\label{ack}
This work is supported by the National Natural Science Foundation of China under
Grants No.~11475189, 11475191, 11135009, and by the 973 Program of China under
Grant No.~2013CB837000, and by the Strategic Priority Research Program
``The Emergence of Cosmological Structures'' of the Chinese
Academy of Sciences, under Grant No.~XDB09000000. W. L. thanks Qiang Yuan, Bin  Yue and Dahai Yan for helpful discussions of parameter constraint, structure formation and extragalactic background light.

\appendix
\section{Halo Mass Function}

The halo mass function ${\rm d}n(z)/{\rm d}M_{\rm vir}$ characterizes the comoving number density distribution of DM halos at different redshifts. It can be usually written in the following widespread formula
\begin{equation}
\frac{{\rm d}n(z)}{{\rm d}M_{\rm vir}}=\frac{\rho_\chi}{M_{\rm vir}}
\sqrt{\frac{2A^2a}{\pi}}\left[1+(a\nu^2)^{-p}\right]\exp\left(-a\nu^2/2
\right)\frac{{\rm d}\nu}{{\rm d}M_{\rm vir}}
\end{equation}
with $(A, \alpha, p) = (0.322, 0.707, 0.3)$, i.e. the well-known Sheth-Tormen formula. $\nu = \delta_c(z)/\sigma(M_{\rm vir})$ and $\delta_c(z) = 1.68[D(z =0)/D(z)]$ is the critical overdensity above which the spherical collapse occurs \citep{1996MNRAS.282..263E}. $D(z)$ is the linear growth factor representing the growth of the density perturbation inside the horizon after matter-radiation equality era. A prevailing approximation can be found in \citep{1992ARA&A..30..499C, 2010gfe..book.....M},
\begin{equation}
\label{eqn:LGFactor}
D(z) \simeq \frac{5\Omega_m/2}{(1+z)[\Omega^{4/7}_m -\Omega_{\Lambda} +(1+\Omega_m/2)(1+\Omega_{\Lambda}/70)]} .
\end{equation}
$\sigma^2(M_{\rm vir})$ is the average variance of the density field, which is evaluated by integrating the matter power spectrum in k-space
\begin{equation}
\sigma^2(M_{\rm vir})=\frac{1}{2\pi^2}\int W^2(kR_{\rm M})P_\delta(k)
k^2{\rm d}k,
\end{equation}
where $W(x)$ is the window function. In the literature, two window functions are often met, i.e. the top-hat window function($W(x)=3(\sin{x}-x \cos{x})/x^3$) and the Gaussian window function($W(x)=\exp[-x^2/2]$). In this paper, we use the former one. $P_{\delta}(k)$ is the matter power spectrum given by
\begin{equation}
P_\delta(k)=A_s(k\cdot{\rm Mpc})^{n_s}T^2(k) .
\end{equation}
In above equation, constant $A_s$ is normalized by $\sigma_8\equiv\sigma(8 h^{-1} \rm Mpc)$. $T(k)$ is the linear transfer function, and here we use its well-fitted form under adiabatic cold DM scenario with $\Omega_{b,0} \ll \Omega_{m,0}$ \citep{1986ApJ...304...15B, 2010gfe..book.....M}
\begin{equation}
T(q)=\frac{\ln(1+2.34q)}{2.34q}\left[1+3.89q+(16.1q)^2+(5.46q)^3+
(6.71q)^4\right]^{-0.25} ,
\end{equation}
where $q=k/\Gamma(h \rm Mpc^{-1})$ and $\Gamma=\Omega_{\rm m,0} h\exp[-\Omega_{\rm b,0}(1+\sqrt{2h}/
\Omega_{\rm m,0})]$ is to describe the horizon scale at $t_{\rm eq}$.

\section{Dark Matter Subhalos in the Galaxy}

When the substructures bring forth, the DM densities with the same radius $r$ are no longer the same. In \citep{2010PhRvD..81d3532K}, the authors defined a probability density function $P(\rho, r)$, which represents at $r$ the probability to take density between $\rho$ and $\rho +\dif \rho$ is $P(\rho, r)~ \dif \rho$. If $f_s$ denotes the fraction of smooth DM component, then $1-f_s$ is that of the clumped component. According to the simulation, $f_s \sim 1$, so the clumpy component only occupies a tiny portion, i.e. $1-f_s \ll 1$. The part of high DM density is postulated to have a power-law distribution. The probability distribution function $P(\rho, r)$ is
\begin{eqnarray}
     P(\rho;r) =& & \frac{f_s}{\sqrt{2\pi \, \Delta^2}} \, \frac{1}{\rho}
     \, \exp \left\{ -\frac{1}{2 \Delta^2} \left[
     \ln\left(\frac{\rho}{\rho_h} e^{\Delta^2/2 }\right)\right]^2 \right\}
      \nonumber \\
     + & &\left(1-f_s\right)
     \frac{1+\alpha(r)}{\rho_h}  \Theta\left(\rho-\rho_h\right)
     \left( \frac{\rho}{\rho_h} \right)^{-(2+\alpha)}.
\label{eqn:pdf}
\end{eqnarray}
The first term comes from the smooth halo component, which has a log-normal distribution with the mean density $\rho_h$ and variance $\Delta^2$. The second term is high-density power-law tail due to substructure. The fraction of smooth-halo part can be well-approximated by
\begin{equation}
\label{eq:fsa}
     f_s(r) = 1 -7\times10^{-3} \left(
     \frac{\bar\rho(r)}{\bar\rho(r=100\,{\rm kpc})} \right)^{-0.26}~,
\end{equation}
where $\bar{\rho}$ is given by the probabilistic average of $\rho$
\begin{eqnarray}
     \bar{\rho}(r) &=& \int_0^{\rho_{\rm max}} \, \rho\,
     P(\rho)\, \dif \rho \nonumber \\
     &=& f_s \rho_h +
 	 (1-f_s)\rho_h \begin{cases}
     \frac{1+\alpha}{\alpha} \left[
     1 - \left(\frac{\rho_{\rm max}}{\rho_h} \right)^{-\alpha} \right];
     & \text{$\alpha\neq0$,} \\
     \ln
     \frac{\rho_{\rm max}}{\rho_h}; & \text{$\alpha=0$}, \end{cases}
\label{eqn:barrho}
\end{eqnarray}
where $\rho_{\rm max} = 80$ GeV cm$^{-3}$. The enhancement due to substructures can be attributed to a boost factor $B(r)$, i.e.
\begin{eqnarray}
     B(r) &=& \frac{ \int\, \rho^2\, dV}{\int\, [\bar\rho(r)]^2\, dV} \nonumber \\
     &=&      \int_{0}^{{\rho_{max}}}
     P(\rho,r)\, \frac{\rho^2}{[\bar\rho(r)]^2}\,d\rho, \nonumber \\
     &=& f_s e^{\Delta^2} +(1-f_s) \frac{1+\alpha}{1-\alpha} \left[\left(\frac{\rho_{max}}{\rho_h} \right)^{1-\alpha} -1 \right].
\label{eqn:standardBF}
\end{eqnarray}
The first term $f_s e^{\Delta^2}$ corresponds to the variation in the smooth component. Since from simulations $\Delta \lesssim 0.2$, it contributes to the overall boost factor by only a few percent and can be safely neglected.

\bibliographystyle{unsrt_update}
\bibliography{ref}
\end{document}